\documentclass{elsart}
\usepackage{natbib,psfig}
\begin{document}
\runauthor{Jarvis \& McLure}
\begin{frontmatter}
\title{The host galaxies of flat-spectrum quasars$^{\star}$}\thanks{Based
    on observations performed at the European Southern Observatory,
    Chile [programme ID: 69.B-0197(A)].}
\author[matt1,matt2]{Matt J.~Jarvis}
\author[ross]{Ross J.~McLure}

\address[matt1]{Sterrewacht Leiden, Postbus 9513, 2300 RA Leiden, The Netherlands}
\address[matt2]{Oxford Astrophysics, Keble Road, Oxford, OX1 3RH, UK}
\address[ross]{IfA, University of Edinburgh, Royal Observatory, Edinburgh EH9 3HJ, UK}

\begin{abstract}
We present the results of deep VLT-ISAAC $K_{s}$-band imaging of four $z
\sim 1.5$ flat-spectrum quasars selected from the Parkes half-Jansky
flat spectrum sample. We find that the hosts of these flat-spectrum
quasars are consistent with lying on the $K-z$ Hubble relation for
radio galaxies. This implies that the flat-spectrum quasar hosts fall in
line with the expectations from orientation based unified schemes and
also that they contain black holes of similar mass. Moreover, the
width of the H$\beta$ broad emission line in these objects tends to be 
narrower than in their misaligned (low-frequency selected quasar)
counterparts, implying that the width of the H$\beta$ broad emission
line depends on source inclination, at least for radio-loud
quasars, in line with previous studies.

\end{abstract}
\end{frontmatter}

\section{Introduction}
It is now widely believed that active galactic nuclei (AGN) are
powered by accretion on to a supermassive black hole. However, it is
still not yet known why some of these active galaxies exhibit powerful
radio emission. The mass of the central black hole appears to play at
least a minor r\^ole in generating powerful ($P_{5 GHz} >
10^{24}$~W~Hz$^{-1}$~sr$^{-1}$) radio emission [e.g. \citet{dun03}], but this is certainly not the only parameter, with
accretion rate and possibly black-hole spin playing crucially
important r\^oles [e.g. see the review by \citet{mcl03}].

It has recently been suggested that the host galaxies of
flat-spectrum, high-frequency selected quasars, inhabit less massive
host galaxies than their low-frequency selected (and therefore
selected independent of orientation) radio-loud quasar counterparts
and also their optically selected counterparts, \citet{osh02}. This in turn leads to the implication that the black holes
in the flat-spectrum quasars (FSQs) fall below the upper envelope in
the radio power -- black-hole mass plane proposed by \citet{dun03}, if the bulge
luminosity -- black-hole mass relation holds out to cosmologically
significant redshifts. This suggests that radio power may not be
correlated with the mass of the supermassive black hole, and must be
due to some other physical process.

\begin{figure}
\psfig{file=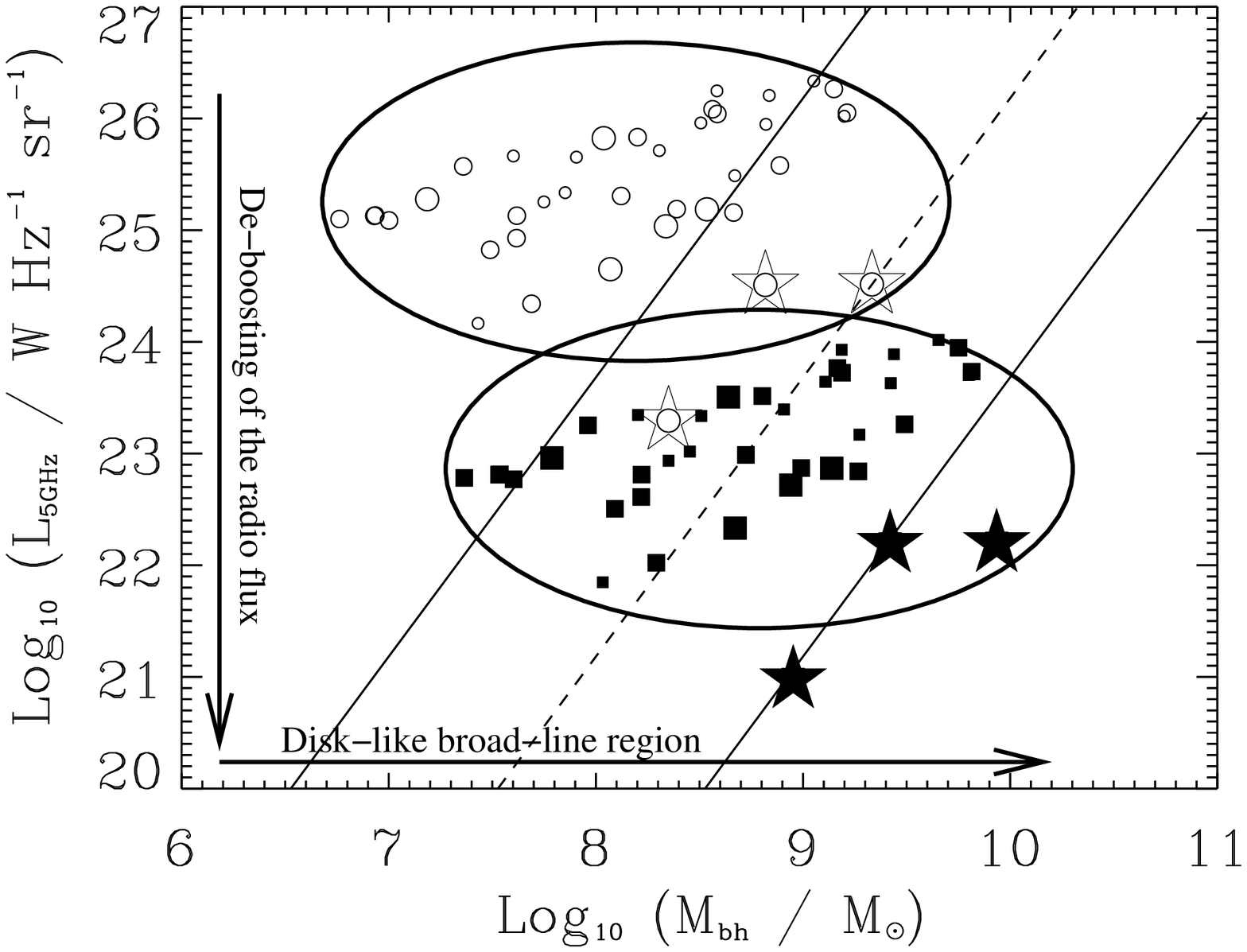,angle=0,width=12.cm}
%\vspace{-0.3cm}
\caption{\label{mjjfig1} The black-hole mass -- radio luminosity plane
adapted from \citet{jm02}, where full details can be found. Open symbols represent points
from the study of \citet{osh02}, the filled symbols represent
where these points would lie after a moderate Doppler boosting
correction and consideration of a disk-like geometry for the
broad-line region. The large stars are anomolous steep-spectrum
objects and are probably not dominated by the core emission.}
\end{figure}

This idea was re-investigated by \citet{jm02} who suggested that the orientation of the
FSQs, with the radio jets preferentially
orientated along the line-of-sight to the observer, may also be a factor. They conclude that
by taking into account the effects of Doppler boosting and also by
assuming the broad-line region may have a disk-like geometry \citep{broth96} that the black-holes in FSQs may be consistent with the upper envelope in black-hole mass (e.g. Fig.~\ref{mjjfig1}).

To investigate this question further we have embarked upon a study of
both the host galaxies and emission-line properties of FSQs selected from the Parkes half-Jansky flat-spectrum sample
\citep{drink97}. In this paper we present some initial
results on the host galaxies of four of these quasars at a redshift
of $z \sim 1.5$.

\section{The Observations}

We have obtained deep $K_{s}$-band observations of seven $z \sim 1.5$
FSQs from the Parkes half-Jansky flat-spectrum sample \citep{drink97} with VLT-ISAAC. The observations were taken in
photometric conditions with typical seeing of 0.5~arcsec. Each object
was observed for a total time of 1.5~hours. A detailed account of the
observations of the complete sample of 7 $z \sim 1.5$ flat-spectrum
quasars will be presented in a forthcoming paper (Jarvis \& McLure in
prep.). The depth of the observations have allowed us to accurately
determine the point-spread function for each source, allowing accurate
modelling and subtraction of the quasar nucleus. This has subsequently
enabled us to determine detailed characteristics of the underlying
host galaxies in these objects.  Furthermore, we have obtained
1.5~hours of $J$-band spectroscopy on each object, again with ISAAC,
to determine the width of the broad H$\beta$ line in each these
objects which should enable us to estimate the mass of the central
supermassive black hole via the virial black-hole mass estimate
\citep{pw00,onken02,mj02}

\section{The host galaxies}
We have performed detailed modelling of the host galaxies of all of the 
FSQs in our sample using the modelling technique described in McLure, 
Dunlop \& Kukula (2000). Full morphological parameters of the
FSQs will be presented in a subsequent paper.

\begin{table}
%\begin{center}
\caption{\label{tab1}Flat-spectrum quasars considered in this paper. Typical error
on the host magnitudes is $\sigma \sim 0.5$~mag.}
\begin{tabular}{lll}
\hline
Source & Redshift & $K_{\rm host}$ \\
\hline
PKS1532+016 & 1.435 &  18.3 \\
PKS1548+056 & 1.422 &  16.9 \\
PKS1602-001 & 1.624 &  18.0 \\
PKS2021-330 & 1.471 &  18.0 \\ 
\hline
\end{tabular}
%\end{center}
\end{table}

\begin{figure}
\psfig{file=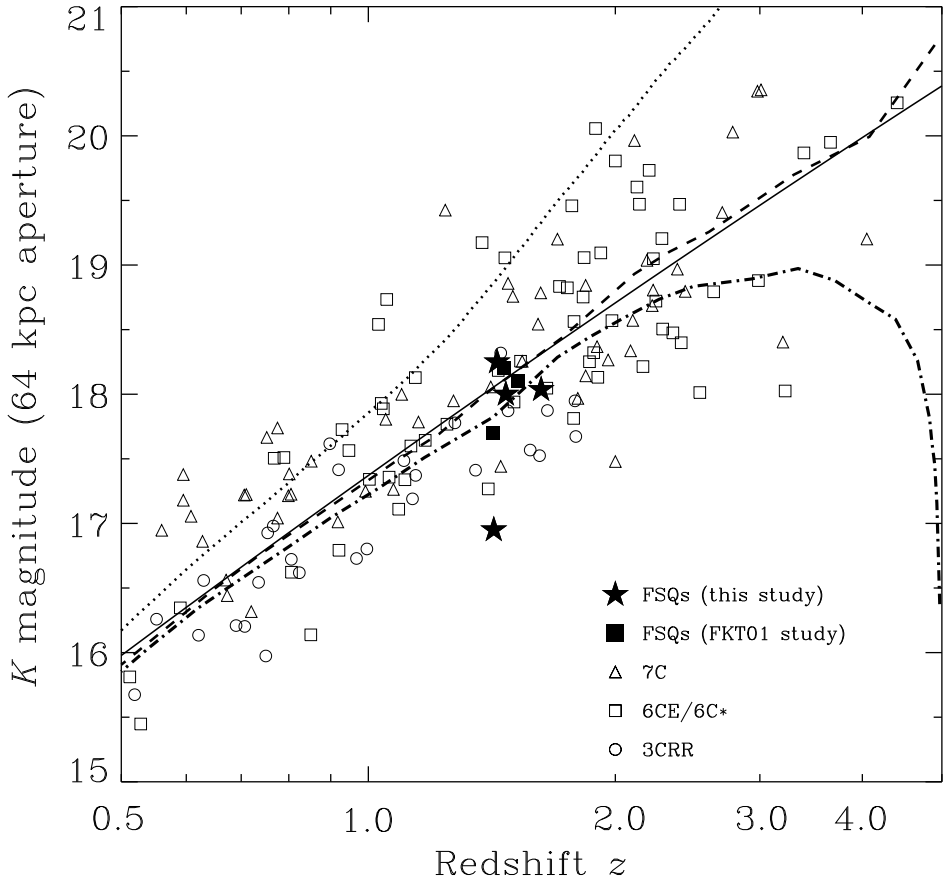,angle=0,width=13.9cm}
%\vspace{-0.3cm}
\caption{\label{kz} The $K - z$ relation for radio galaxies from
  low-frequency selected radio samples. 3CRR (open circles, Laing,
  Riley \& Longair 1983), 6CE/6C* (open squares, Eales et al. 1997; Jarvis et al. 2001), 7C (open triangles, Willott et al. 2003; Lacy et
  al. 2000), the Parkes flat-spectrum quasars from this study (filled
  stars) and those from the study of Falomo, Kotilainen \& Treves
  (2001; filled squares, adjusted to $K$ magnitude using $H-K = 0.5$
  at $z \sim 1.5$). The curves represent various evolutionary
  models for a 3~$L_{\star}$ galaxy (see Willott et al. (2003) for
  full details). }
\end{figure}

In this paper we focus on the $K$-band magnitudes determined for the
host galaxies and compare these with the radio galaxy host magnitudes
which are uncontaminated by the bright point source resulting from the
quasar nucleus. Table~\ref{tab1} lists the sources with there relevant
properties, Fig.~\ref{kz} shows that all of these sources are
consistent with lying on the radio galaxy $K-z$ Hubble diagram
\citep{eales97,jar01,cjw03} implying that the hosts of these powerful
radio selected quasars are some of the most massive galaxies in the
Universe, i.e. $> 3~L_{\star}$.

This is in
apparent opposition to the conclusions reached by \citet{osh02}. From their analysis of lower redshift flat-spectrum
radio-loud quasars they suggested that the radio-loud quasars possibly
reside in less luminous galaxies in comparison to their optically
selected quasar counterparts. Although this may well be true to some
degree with respect to the optical selection biases, we find that
genuinely powerful radio-loud AGN (i.e. $P_{5 \rm{GHz}} >
10^{24}$~W~Hz$^{-1}$~sr$^{-1}$) do seem to inhabit super-$L^{\star}$
galaxies over all cosmic epochs.

\section{Black hole masses}
In this section we utilize the correlation between black-mass and
bulge luminosity ($M_{\rm bh} - L_{\rm bulge}$) to estimate the
black-hole masses in these FSQs.

Using the relation between black-hole mass and bulge luminosity given
in McLure \& Dunlop (2002), i.e.
\begin{displaymath}
\log(M_{\rm bh}/M_{\odot}) = -0.50(\pm 0.02)M_{R} - 2.96(\pm0.48),
\end{displaymath}

we can estimate the mass of the central supermassive black holes in
these quasars.
Assuming that the host galaxies evolve along the tracks shown in
Fig.~\ref{kz} then the host galaxies of the FSQs are
all larger than 3~$L_{\star}$ (using $M_{K}^{\star} = -23.52$ from
Kochanek et al. 2001). Using $R-K = 2.7$ for local ellipticals we find
that according to the $M_{\rm bh} - L_{\rm bulge}$ relation that all
of our FSQs contain black-holes with $10^{8}~{\rm M}_{\odot}
< M_{\rm bh} < 10^{9}~{\rm M}_{\odot}$. 

This is in line with the upper envelope in the black-hole mass --
radio power plane proposed by Dunlop et al. (2003), where genuinely
powerful radio-loud sources are powered by black holes with mass $>
10^{8}{\rm M}_{\odot}$.

This implies that targeting {\it powerful} radio-loud quasars or
radio galaxies allows one to probe the upper bounds in black-hole mass
at any given epoch. 

\section{The virial black-hole mass estimator: a word of caution}

In this section we briefly discuss the possible orientation bias in
using the virial black-hole mass estimator. It has been shown that the
width of the broad emission lines in radio-loud quasars are dependent
on the orientation of the source (e.g. Wills \& Browne 1986;
Brotherton 1996; Vestergaard, Wilkes \& Barthel 2000), with the more
pole-on sources, i.e. those with high radio core-to-lobe flux ratios,
having narrower broad lines. Therefore,
according to most forms of the orientation based Unified Schemes for
active galaxies, the flat-spectrum quasars, which are presumed to be pole-on powerful radio-loud quasars, should exhibit the
narrowest broad lines. 

The FSQs from the sample of Oshlack et al. (2002)
have a mean FWHM of the H$\beta$ broad emission line of $\sim
3500$~km~s$^{-1}$. On the contrary, the low-frequency selected quasars
from the Molonglo Quasar Sample (Baker et al. 1999), where the low-frequency gives an
orientation independent selection technique, has a mean FWHM(H$\beta$)
$\sim 7000$~km~s$^{-1}$. This at least implies that orientation may
need to be considered when using the virial black-hole mass estimator,
at least for radio-loud quasars. In a subsequent paper we investigate
this further for our sample of $z \sim 1.5$ FSQs with
deep $J-$band spectroscopy around the H$\beta$ emission line.

\section*{Ackowledgements}
MJJ thanks the Lorentz centre at Leiden University for their excellent
support in organising this meeting, and in particular Gerda Filippo,
without whom this meeting would never have run so smoothly and also
Huub R\"ottgering and George Miley who found the extra money so that I could
stay in Leiden for the extra two months which encompassed this
meeting.

\end{document}